\documentclass[letterpaper, 10 pt, conference]{ieeeconf}
\usepackage{amsmath}
\usepackage{amssymb}
\usepackage{balance}
\usepackage{float}
\usepackage{graphicx}
\usepackage{mathtools}
\usepackage[hidelinks]{hyperref}
\usepackage{algpseudocode}
\usepackage[linesnumbered,ruled,vlined]{algorithm2e}
\usepackage[sorting=none]{biblatex}
\IEEEoverridecommandlockouts
\SetKwInput{KwInput}{Input}
\SetKwInput{KwOutput}{Output}

\title{Data-Driven Estimation of Quadrotor Motor Efficiency via Residual Minimization}
\author{
Sheng-Wen Cheng$^{1}$ \thanks{$^{1}$The University of Texas at Austin, Department of Computer Science, USA, E-mail: \texttt{shengwen.c@utexas.edu}.} and
Teng-Hu Cheng$^{2}$ \thanks{$^{2}$National Yang Ming Chiao Tung University, Department of Mechanical Engineering, Hsinchu 30010, Taiwan, E-mail: \texttt{tenghu@nycu.edu.tw}.}
}

\begin{document}

\maketitle

\begin{abstract}
A data-driven framework is proposed for online estimation of quadrotor motor efficiency via residual minimization. The problem is formulated as a constrained nonlinear optimization that minimizes trajectory residuals between measured flight data and predictions generated by a quadrotor dynamics model. A sliding-window strategy enables online estimation, and the optimization is efficiently solved using an iteratively reweighted least squares (IRLS) scheme combined with a primal-dual interior-point method, with inequality constraints enforced through a logarithmic barrier function. Robust z-score weighting is employed to reject outliers, which is particularly effective in motor clipping scenarios where the proposed estimator exhibits smaller spikes than an EKF baseline. Compared to traditional filter-based approaches, the batch-mode formulation allows selective inclusion of data segments via IRLS reweighting and hard-rejection. This structure is well-suited for online estimation and supports applications such as fault detection and isolation (FDI), health monitoring, and predictive maintenance in aerial robotic systems. Simulation results under various degradation scenarios demonstrate the accuracy and robustness of the proposed estimator.
\end{abstract}

\section{Introduction}

A quadrotor is a highly agile aerial robotic platform that relies on rapid motor response for stable and precise flight. The performance of its propulsion system directly impacts energy efficiency, flight time, and maneuverability. Motor efficiency, a critical yet often unmeasured factor, can deteriorate due to elevated temperatures, component aging, mechanical wear, and battery voltage fluctuations. An effective estimator holds significant potential for fault detection and isolation (FDI), early failure prediction, and maintenance alerts, ultimately mitigating in-flight failure risks.

Extensive research on system identification for quadrotors has been conducted to improve control accuracy and state estimation. Most prior works focus on identifying parameters such as mass, inertia, or center of mass, and detecting actuator faults. For example, \cite{svacha2020imu} employed an IMU combined with motor speed commands and an Unscented Kalman Filter (UKF) to estimate mass and inertia using Newton-Euler dynamics, while \cite{kwon2023real} used a Kalman Filter (KF) combined with Recursive Least Squares (RLS) to estimate the moment of inertia. \cite{abas2011parameter} utilized UKF for estimating inertia, velocity, and angular velocity, and \cite{ho2017mass} compared least-squares, EKF, and instrumental variable methods for mass estimation. Other approaches include \cite{lei2024robust}, who proposed a robust geometric control framework for quadrotors that incorporates online estimation of inertia parameters, and \cite{wuest2019online} employed UKF and EKF to estimate geometric and inertial parameters. \cite{das2025dronediffusion} used diffusion models to learn unmodeled effects such as aerodynamic disturbances, mass uncertainty, and sensor noise. For fault detection, methods such as neural networks \cite{garg2023neural}, augmented-state Kalman filters \cite{zhong2018robust}, sparse identification and Thau’s observer \cite{lee2024data}, and nonlinear observers with sliding mode control \cite{ahmadi2023active} have been proposed. \cite{eschmann2024data} further demonstrated a data-driven system identification framework using Maximum A Posteriori (MAP) estimation to jointly estimate motor delays and inertial properties. However, relatively little attention has been given to the estimation of motor efficiency for quadrotors, despite its critical impact on energy consumption, stability, and flight time. Among these studies, our previous work \cite{chang2023motor} introduced a UKF-based estimator integrated into a cooperative multirotor system.

This work proposes a batch-mode, optimization-based framework for estimating quadrotor motor efficiency by minimizing trajectory residuals. While not explored in this study, such formulations also offer potential for integration with learning-based methods. The estimation problem is posed as a least-squares optimization with inequality constraints to ensure the efficiency factors remain within a physically valid range (typically $0 \leq \eta \leq 1$). To solve this constrained optimization problem, a primal-dual interior-point method is adopted, augmented with an outer IRLS loop that applies robust z-score weighting to suppress outliers. The inequality constraints are enforced through a logarithmic barrier function, and the interior-point framework was originally introduced by Karmarkar \cite{karmarkar1984new} and later extended to nonlinear problems using barrier methods by Fiacco and McCormick \cite{fiacco1990nonlinear}. A comprehensive introduction to interior-point techniques, including log-barrier formulations, can be found in \cite{boyd2004convex} by Boyd and Vandenberghe. The main contributions of this work are summarized as follows:

\begin{itemize}
  \item[$\bullet$] A nonlinear optimization-based estimator with bounded constraints is developed to identify quadrotor motor efficiency by minimizing residuals between measured and predicted trajectories.
  \item[$\bullet$] A sliding-window scheme with a primal-dual interior-point method and IRLS reweighting improves robustness by rejecting outliers and reducing spikes.
  \item[$\bullet$] The proposed approach is validated in simulation under diverse motor degradation and fault scenarios.
\end{itemize}

The rest of this paper is organized as follows. Section \ref{sec:quadrotor} describes the quadrotor dynamics and control framework. Section \ref{sec:estimator} presents the proposed motor efficiency estimator. Section \ref{sec:simulation} provides the simulation results. Finally, Section \ref{sec:conclusion} concludes the paper.

\section{Modeling and Control of Quadrotor}
\label{sec:quadrotor}

\subsection{Quadrotor Dynamics}

The quadrotor rigid-body dynamics are divided into translational and rotational components, capturing the effects of thrust, gravity, and inertia. The equations of motion are:
\begin{align}
\dot{x} &= v \label{eq:xv} \\
m \dot{v} &= mge_3 - f_{\text{c}} Re_3 \label{eq:mv} \\
\dot{R} &= R [\Omega]_{\times} \label{eq:Rdot} \\
J\dot{\Omega} + \Omega \times J \Omega &= M, \label{eq:JW_WJW_M}
\end{align}

\noindent where $x\in \mathbb{R}^3$ is the position, $v \in \mathbb{R}^3$ is the velocity, $R \in \text{SO}(3)$ is the rotation matrix from the body-fixed frame to the inertial frame, $\Omega \in \mathbb{R}^3$ is the angular velocity, $M\in \mathbb{R}^3$ is the moment, $J \in \mathbb{R}^{3\times3}$ is the inertia matrix, $f_{\text{c}}\in \mathbb{R}$ is the collective thrust, $g\in \mathbb{R}$ is the gravitational acceleration, and $e_3 = [0,0,1]^{\top}$ is the unit vector along the inertial z-axis. The operator $[\cdot]_{\times}$ maps a vector to a skew-symmetric matrix $\mathbb{R}^{3} \mapsto \mathfrak{so}(3)$:
\begin{equation}
\begin{aligned}
[\Omega]_{\times} \coloneqq
\begin{bmatrix}
0 & -\Omega_3 & \Omega_2 \\
\Omega_3 & 0 & -\Omega_1 \\
-\Omega_2 & \Omega_1 & 0
\end{bmatrix}.
 \label{eq:skew_matrix}
\end{aligned}
\end{equation}

\noindent For simplicity, the inertia matrix in \eqref{eq:JW_WJW_M} is assumed diagonal:
\begin{equation}
J = \operatorname{diag}(J_{xx}, J_{yy}, J_{zz}). \notag
\end{equation}

\subsection{Geometric Tracking Control of Quadrotor}

Following the geometric tracking control on $\text{SE}(3)$ developed by Lee et al. \cite{lee2010geometric}, the controller is designed to track desired trajectories. The tracking errors are defined as:
\begin{align}
e_x &= x - x_d \label{eq:ex} \\
e_v &= v - v_d \\
e_R &= \frac{1}{2}(R_d^{\top} R - R^{\top} R_d)^{\vee} \\
e_{\Omega} &= \Omega - R^{\top} R_d \Omega_d, \label{eq:eW}
\end{align}

\noindent where $(\cdot)^\vee$ denotes the inverse of \eqref{eq:skew_matrix}, i.e., $\mathfrak{so}(3) \mapsto \mathbb{R}^{3}$, and $(x_d,v_d,R_d,\Omega_d)$ are the desired position, velocity, attitude, and angular velocity. Based on \eqref{eq:ex}--\eqref{eq:eW}, the control inputs of thrust $f_{\text{c}}$ and moment $M$ are designed as:
\begin{align}
f_{\text{c}} =& -(-k_x e_x - k_v e_v - mge_3 + m \ddot{x}_d) \cdot R e_3 \label{eq:fc} \\
M = &-k_R e_R - k_{\Omega} e_{\Omega} + \Omega \times J \Omega \notag \\
&-J ([\Omega]_{\times} R^{\top}R_d \Omega_d - R^{\top} R_d \dot{\Omega}_d), \label{eq:M}
\end{align}

\noindent where $k_x$, $k_v$, $k_R$, $k_{\Omega}$, are control gains.

\subsection{Thrust Allocation and Motor Efficiency Modeling}

The mapping between individual motor thrusts and the collective force and moments acting on the quadrotor can be represented by a linear transformation. Define the motor thrust vector:
\begin{equation}
\begin{aligned}
f_{\text{m}}=\begin{bmatrix}
f_1 &
f_2 &
f_3 &
f_4
\end{bmatrix}^{\top},
\label{eq:fm}
\end{aligned}
\end{equation}

\noindent where $f_i \in \mathbb{R}$ denotes the thrust by motor $i$ in \eqref{eq:fm}. The actual collective thrust $f_{\text{c}}$ and moment $M=[M_1,M_2,M_3]^{\top}$ are related to the motor thrusts, scaled by an efficiency factor, with respect to the control inputs computed in \eqref{eq:fc}--\eqref{eq:M}:
\begin{equation}
\begin{aligned}
\begin{bmatrix}
f_{\text{c},\text{actual}} \\ M_{1,\text{actual}} \\ M_{2,\text{actual}} \\ M_{3,\text{actual}}
\end{bmatrix} = \Lambda Ef_{\text{m}} = \Lambda E\Lambda^{-1}
\begin{bmatrix}
f_{\text{c}} \\ M_{1} \\ M_{2} \\ M_{3}
\end{bmatrix},
\end{aligned}
\label{eq:LambdaEfm}
\end{equation}

\noindent where $\Lambda$ in \eqref{eq:LambdaEfm} is the thrust allocation matrix, and $E$ is a diagonal matrix representing the motor efficiency. The thrust allocation matrix $\Lambda$ is defined as:
\begin{equation}
\begin{aligned}
\Lambda=\begin{bmatrix}
1 & 1 & 1 & 1 \\
-d & d & d & -d \\
d & d & -d & -d \\
-c_{\tau f} & c_{\tau f} & -c_{\tau f} & c_{\tau f}
\end{bmatrix},
\end{aligned}
\label{eq:Lambda}
\end{equation}

\noindent where $d \in \mathbb{R}$ is the distance parameter from the center of mass, and $c_{\tau f} \in \mathbb{R}$ is the thrust-to-torque coefficient in \eqref{eq:Lambda}. The motor efficiency matrix $E$ is given by:
\begin{equation}
\begin{aligned}
E = \operatorname{diag}(\eta_1, \eta_2, \eta_3, \eta_4). \label{eq:E}
\end{aligned}
\end{equation}

\noindent In \eqref{eq:E}, $\eta_i$ denotes the efficiency of motor $i$. Assuming perfect efficiency ($\eta_i = 1, \forall i$), the control commands $f_{\text{c}}$ and $M$ in \eqref{eq:fc}--\eqref{eq:M} exactly match the actual $f_{\text{c, \text{actual}}}$ and $M_{\text{actual}}$ generated by the motors in \eqref{eq:LambdaEfm}.

\subsection{Numerical Integration of Quadrotor Dynamics}

This section presents the numerical integration to predict states for motor efficiency estimation. The updates use the quadrotor dynamics model in \eqref{eq:xv}--\eqref{eq:JW_WJW_M}, combined with the actual thrust and moment model computed in \eqref{eq:LambdaEfm}. The predicted states are later compared to measurements to construct trajectory residuals. Define the motor efficiency vector as:
\begin{equation}
\begin{aligned}
s =
\begin{bmatrix}
\eta_1 &
\eta_2 &
\eta_3 &
\eta_4
\end{bmatrix}^{\top}.
\end{aligned}
\label{eq:s}
\end{equation}

\noindent The velocity is updated by integrating the translational acceleration over a small time interval $\Delta t \in \mathbb{R}$:
\begin{align}
\hat{v}_{t+1}(s) = v_t + \bigg (ge_3 - \frac{f_{\text{c},\text{actual}}Re_3}{m} \bigg) \Delta t.
\label{eq:v_integrate}
\end{align}

\noindent The position is updated by integrating the velocity and acceleration:
\begin{align}
\hat{x}_{t+1}(s) = x_t + v_t \Delta t + \frac{1}{2} \bigg (ge_3 - \frac{f_{\text{c},\text{actual}}Re_3}{m} \bigg) \Delta t^2.
\label{eq:x_integrate}
\end{align}

\noindent The angular velocity is integrated from the rotational dynamics:
\begin{align}
\hat{\Omega}_{t+1}(s) = \Omega_t + \big[J^{-1}(M_{\text{actual}} - \Omega \times J \Omega) \big]\Delta t.
\label{eq:W_integrate}
\end{align}

\noindent For the rotation matrix update, a first-order approximation is used:
\begin{equation}
\begin{aligned}
\hat{R}_{t+1}(s) = R_t (I + [\hat{\Omega}_t(s)]_{\times} \Delta t).
\end{aligned}
\label{eq:R_integrate}
\end{equation}

\section{Motor Efficiency Estimation}
\label{sec:estimator}

\subsection{Overview of the Estimation Framework}

\begin{figure}[h]
    \centering
    \includegraphics[width=0.99\linewidth]{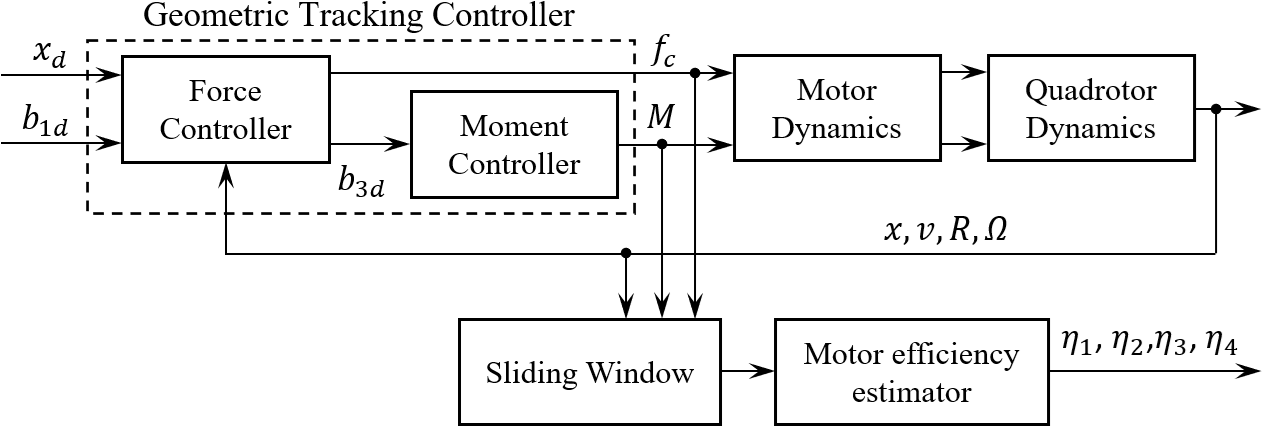}
    \caption{Architecture of the proposed motor efficiency estimator integrated with geometric tracking controller.}
    \label{fig:block_diagram}
\end{figure}

Fig. \ref{fig:block_diagram} illustrates the proposed motor efficiency estimation framework. The estimator operates alongside the geometric tracking controller and takes as input the desired thrust and motor commands computed by the controller. Within each sliding window of length $n$, the system collects the measured states including position $x$, velocity $v$, angular velocity $\Omega$, and rotation matrix $R$, together with the control inputs of collective thrust $f_c$ and moment $M$. These measurements are compared with model-based predictions, and the resulting residuals are minimized through constrained nonlinear optimization to estimate the efficiency of each motor.

\subsection{Definition of Trajectory Residuals}

Trajectory residuals quantify the discrepancy between measured and predicted states from \eqref{eq:v_integrate}--\eqref{eq:R_integrate} with estimated efficiency vector $s$ in \eqref{eq:s}. The position, velocity, and angular velocity residuals are defined as:
\begin{align}
r_{v,t}(s) &= v_t - \hat{v}_t(s) \label{eq:rv} \\
r_{x,t}(s) &= x_t - \hat{x}_t(s) \label{eq:rx} \\
r_{\Omega,t}(s) &= \Omega_t - \hat{\Omega}_t(s). \label{eq:rW}
\end{align}

\noindent To define the orientation residual, the relative rotation between two consecutive time steps is extracted by:
\begin{align}
R_{t+1} &= R_t \delta R  \notag \\
\rightarrow \delta R &= R_t^{\top} R_{t+1}.
\label{eq:deltaR_measured}
\end{align}

\noindent On the other hand, assuming small $\Delta t$, the predicted incremental rotation is approximated by:
\begin{align}
\delta \hat{R}(s) &=\exp([\hat{\Omega}_t(s)]_{\times} \Delta t) \notag \\ &\approx I + [\hat{\Omega}_t(s)]_{\times} \Delta t.
\label{eq:deltaR_predicted}
\end{align}

\noindent With \eqref{eq:deltaR_measured} and \eqref{eq:deltaR_predicted}, the orientation residual is defined as:
\begin{align}
r_{R,t}(s) &= \frac{1}{2} \operatorname{tr}[I - \delta R^{\top} \delta \hat{R}(s)]. \label{eq:rR}
\end{align}

\noindent Stacking \eqref{eq:rv}--\eqref{eq:rW} and \eqref{eq:rR} over a sliding window of $n$-segments gives:
\begin{equation}
\begin{aligned}
r(s) = \frac{1}{\sqrt{n}}
\begin{bmatrix}
r_{v,t=0}(s) \\
r_{x,t=0}(s) \\
r_{\Omega,t=0}(s) \\
r_{R,t=0}(s) \\
\vdots \\
r_{v,t=n-1}(s) \\
r_{x,t=n-1}(s) \\
r_{\Omega,t=n-1}(s) \\
r_{R,t=n-1}(s)
\end{bmatrix}.
\end{aligned}
\label{eq:r_full}
\end{equation}

\noindent The aggregated residual vector in \eqref{eq:r_full} is used to estimate motor efficiency factors via optimization.

\subsection{Formulation of the Optimization Problem}

The motor efficiency estimation is formulated as a constrained nonlinear optimization problem. The objective function comprises two terms: (i) a trajectory residual penalty term and (ii) a temporal smoothness regularization term that promotes temporal smoothness across successive estimates. The cost function at time $t$ is defined as:
\begin{equation}
\begin{aligned}
F(s_t) = \frac{1}{2} \big \Vert r(s_t) \big \Vert^2_{G} + \frac{\gamma}{2} \big \Vert s_{t} - s_{t-1} \big \Vert^2,
\end{aligned}
\label{eq:F_cost}
\end{equation}

\noindent where $\Vert r(s_t) \Vert_{G} = \sqrt{r(s_t)^{\top} G r(s_t)}$ denotes the weighted norm of the residual vector $r(s_t)$, and $G \in \mathbb{R}^{10n\times10n}$ is a positive semidefinite block-diagonal weighting matrix. The local weight matrix for each time segment is:
\begin{align}
G_i =
\operatorname{diag}
(&g_{v_1},
g_{v_2},
g_{v_3},
g_{x_1},
g_{x_2},
g_{x_3}, \notag \\
& g_{\Omega_1},
g_{\Omega_2},
g_{\Omega_3},
g_{R}).
\label{eq:Wi}
\end{align}

\noindent Using \eqref{eq:Wi}, the full weighting matrix is assembled as:
\begin{equation}
\begin{aligned}
G = \operatorname{blockdiag} (w_0 G_0, \cdots, w_{n-1} G_{n-1}),
\end{aligned}
\label{eq:W}
\end{equation}

\noindent where $w_i$ is the weight for outlier rejection that will be introduced in the next section. Finally, with the objective function \eqref{eq:F_cost} with $G$-norm defined in \eqref{eq:W}, the optimization problem is defined as:
\begin{equation}
\begin{aligned}
\min_{s_t} \quad &  F(s_t) \\
\textrm{s.t.} \quad & \eta_{\text{min}} \leq \eta_i \leq \eta_{\text{max}} ,\\ & i = 1, \cdots, 4.
\label{eq:optimization_problem}
\end{aligned}
\end{equation}

\noindent These constraints ensure that each motor's efficiency remains within physically valid bounds.

\subsection{Outlier Rejection and Reweighting}

To suppress the influence of abnormal measurements, a median absolute deviation (MAD)–based reweighting scheme is applied at each outer iteration. MAD is used as a robust scale estimator because extreme residuals can bias the standard deviation estimate. For the $i$-th segment, the weighted residual energy is
\begin{equation}
\begin{aligned}
e_i = r_i(s)^{\top} G_i r_i(s).
\end{aligned}
\label{eq:residual_energy}
\end{equation}

\noindent The residual energy median and MAD are computed as
\begin{equation}
\begin{aligned}
\tilde{m} &= \operatorname{median}(e_i), \\
\text{MAD} &= 1.4826 \cdot \operatorname{median}(\vert e_i - \tilde{m} \vert)
\end{aligned}
\end{equation}

\noindent Each residual energy is normalized to a robust z-score:
\begin{equation}
\begin{aligned}
z_i = \frac{\vert e_i - \tilde{m} \vert}{\max(\text{MAD}, \epsilon_{\text{min}})},
\end{aligned}
\end{equation}

\noindent where $\epsilon_{\text{min}} > 0$ avoids division by zero, and the corresponding weight is updated by
\begin{equation}
\begin{aligned}
w_i = \max \bigg (\frac{1}{1+(\frac{z_i}{z_\text{soft}})^p}, w_{\text{min}} \bigg),
\end{aligned}
\end{equation}
where $z_\text{soft}$ controls the decay threshold, $p$ specifies the decay rate, and $w_{\text{min}}$ denotes the minimum weight. To further discard extreme outliers, a hard rejection is imposed:
\begin{align}
 w_i =
\begin{cases} 
 0       & \text{if\ } z_i > z_{\text{hard}} \\
 w_i & \text{otherwise}.
\end{cases}
\label{eq:hard_reject}
\end{align}

\noindent \eqref{eq:residual_energy}--\eqref{eq:hard_reject} jointly apply soft decay and hard rejection, which attenuate moderate outliers while eliminating extreme ones, thereby improving robustness to noise and modeling errors.

\subsection{KKT Conditions for Constrained Optimization}

To solve the constrained optimization problem of \eqref{eq:optimization_problem}, a primal-dual interior-point method is adopted. The inequality constraints are reformulated in the standard form:
\begin{equation}
\begin{aligned}
\phi(s_t) \preceq 0,
\end{aligned}
\label{eq:constraint_standard}
\end{equation}

\noindent where \eqref{eq:constraint_standard} is defined as:
\begin{equation}
\begin{aligned}
\phi(s) =
\begin{bmatrix}
\phi_1(s) \\
\phi_2(s) \\
\phi_3(s) \\
\phi_4(s) \\
\phi_5(s) \\
\phi_6(s) \\
\phi_7(s) \\
\phi_8(s) \\
\end{bmatrix}
= \begin{bmatrix}
\eta_1 - \eta_{\text{max}} \\
\eta_2 - \eta_{\text{max}} \\
\eta_3 - \eta_{\text{max}} \\
\eta_4 - \eta_{\text{max}} \\
\eta_{\text{min}} - \eta_1 \\
\eta_{\text{min}} - \eta_2 \\
\eta_{\text{min}} - \eta_3 \\
\eta_{\text{min}} - \eta_4
\end{bmatrix}.
\end{aligned}
\label{eq:phi}
\end{equation}

\noindent A corresponding dual variable $\lambda \in \mathbb{R}^m$, $m=8$ is introduced, associated with the inequality constraints defined by \eqref{eq:constraint_standard} and \eqref{eq:phi}. The Karush–Kuhn–Tucker (KKT) conditions for optimality consist of the following:

\noindent $\bullet$ \textbf{Stationarity:}
\begin{equation}
\begin{aligned}
\nabla F(s_t) + \sum_{i=1}^{m} \lambda_i \nabla \phi_i(s_t) = 0.
\end{aligned}
\label{eq:kkt1}
\end{equation}

\noindent $\bullet$ \textbf{Primal feasibility:}
\begin{equation}
\begin{aligned}
\phi_i(s_t) &\leq 0,\ i = 1, \cdots, m.
\end{aligned}
\label{eq:kkt2}
\end{equation}

\noindent $\bullet$ \textbf{Dual feasibility:}
\begin{equation}
\begin{aligned}
\lambda_i & \geq 0,\ i = 1, \cdots, m.
\end{aligned}
\label{eq:kkt3}
\end{equation}

\noindent $\bullet$ \textbf{Complementary slackness:}
\begin{equation}
\begin{aligned}
\lambda_i & \phi_i(s_t) = 0,\ i = 1, \cdots, m.
\end{aligned}
\label{eq:kkt4}
\end{equation}

\noindent The conditions of \eqref{eq:kkt1}--\eqref{eq:kkt4} are enforced iteratively using a primal-dual interior-point method, which ensures convergence to a solution satisfying optimality and feasibility. Details of the solution strategy are provided in the following section.

\subsection{Primal-Dual Interior-Point Method}

To handle the inequality-constrained optimization in \eqref{eq:optimization_problem}, the problem is reformulated using a logarithmic barrier function:
\begin{gather}
\min_{s_t} \quad  F(s_t) - \frac{1}{\beta} \sum_{i=1}^{8} \log\big (-\phi_i(s_t) \big).
\label{eq:log_cost}
\end{gather}

\noindent The stationarity condition of \eqref{eq:log_cost} is derived as:
\begin{equation}
\begin{aligned}
\nabla F(s_t) + \frac{1}{\beta} \sum_{i=1}^{8} \bigg ( \frac{1}{-\phi_i(s_t)}  \nabla \phi_i(s_t) \bigg ) = 0.
\end{aligned}
\label{eq:lambda}
\end{equation}

\noindent By observing \eqref{eq:lambda}, the dual variables $\lambda_i$ can be expressed as:
\begin{equation}
\begin{aligned}
\lambda_i = \frac{1}{-\beta \phi_i(s_t)}.
\end{aligned}
\label{eq:dual}
\end{equation}

\noindent Rearranging \eqref{eq:dual} yields:
\begin{equation}
\begin{aligned}
-\lambda_i \phi_i(s_t) - \frac{1}{\beta} = 0.
\end{aligned}
\label{eq:modified_cs}
\end{equation}

\noindent Equation \eqref{eq:modified_cs} is called the modified complementary slackness condition, which approaches the standard complementary slackness condition \eqref{eq:kkt4} as $\beta \rightarrow \infty$, forming the central path followed by the primal-dual interior-point method. The KKT residual vector is defined as:
\begin{align}
& r_{\text{KKT}}(s, \lambda) =
\begin{bmatrix}
\nabla F(s_t) + D\phi(s_t)^{\top} \lambda \\
-\operatorname{diag}(\lambda) \phi(s_t) - (1/\beta)\mathbf{1}
\end{bmatrix} \notag
\\ &=
\begin{bmatrix}
J_r(s_{t})^{\top} G r(s_{t}) + \gamma (s_t - s_{t-1}) + D\phi(s_t)^{\top} \lambda \\
-\operatorname{diag}(\lambda) \phi(s_t) - (1/\beta)\mathbf{1}
\end{bmatrix},
\label{eq:kkt_residual}
\end{align}

\noindent where $D\phi(s)$ is the Jacobian matrix of constraints in \eqref{eq:kkt_residual}:
\begin{equation}
\begin{aligned}
D\phi(s) =
\begin{bmatrix}
\nabla \phi_1(s)^{\top} \\
\nabla \phi_2(s)^{\top} \\
\nabla \phi_3(s)^{\top} \\
\nabla \phi_4(s)^{\top} \\
\nabla \phi_5(s)^{\top} \\
\nabla \phi_6(s)^{\top} \\
\nabla \phi_7(s)^{\top} \\
\nabla \phi_8(s)^{\top} \\
\end{bmatrix}
= \begin{bmatrix}
1 & 0 & 0 & 0 \\
0 & 1 & 0 & 0 \\
0 & 0 & 1 & 0 \\
0 & 0 & 0 & 1 \\
-1 & 0 & 0 & 0 \\
0 & -1 & 0 & 0 \\
0 & 0 & -1 & 0 \\
0 & 0 & 0 & -1
\end{bmatrix}.
\label{eq:Dphi}
\end{aligned}
\end{equation}

\noindent Along with \eqref{eq:Dphi}, the Jacobian of trajectory residuals $J_r(s)$ in \eqref{eq:kkt_residual} is constructed by stacking residual derivatives across $n$ segments:
\begin{equation}
\begin{aligned}
J_r(s) =
\begin{bmatrix}
J_{r, 0} (s) \\
\vdots \\
J_{r, n-1} (s)
\end{bmatrix},\ 
J_{r, i} (s) =
\begin{bmatrix}
\frac{\partial r_{v,t}(s)}{\partial s} \\
\frac{\partial r_{x,t}(s)}{\partial s} \\
\frac{\partial r_{\Omega,t}(s)}{\partial s} \\
\frac{\partial r_{R,t}(s)}{\partial s}
\end{bmatrix}.
\end{aligned}
\label{eq:Jr}
\end{equation}

\noindent Defining primal-dual variable $y = [s, \lambda]^{\top}$ and applying Newton's method to linearize the KKT residuals:
\begin{equation}
\begin{aligned}
r_{\text{KKT}}(y + \Delta y) \approx r_{\text{KKT}}(y) + Dr_{\text{KKT}}(y) \Delta y = 0,
\end{aligned}
\label{eq:kkt_linearized}
\end{equation}

\noindent which leads to the following linear system:
\begin{equation}
\begin{aligned}
\underbrace{
\begin{bmatrix}
J_r(s_{t})^{\top} G J_r(s_{t}) + \gamma I & D\phi(s_t)^{\top} \\
-\operatorname{diag}(\lambda) D \phi(s_t) & -\operatorname{diag}(\phi(s_t))
\end{bmatrix}
}_{Dr_{\text{KKT}}(y)}
\underbrace{
\begin{bmatrix}
\Delta s_t \\
\Delta \lambda
\end{bmatrix}
}_{\Delta y}
= -
\underbrace{
\begin{bmatrix}
r_{\text{dual}} \\
r_{\text{cent}}
\end{bmatrix}
}_{r_{\text{KKT}}(y)}.
\label{eq:kkt_system}
\end{aligned}
\end{equation}

\noindent Solving the Newton system \eqref{eq:kkt_system} gives the primal-dual update direction:
\begin{equation}
\begin{aligned}
\Delta y = - (Dr_{\text{KKT}}(y))^{-1} r_{\text{KKT}}(y).
\end{aligned}
\label{eq:delta_y}
\end{equation}

\noindent To preserve feasibility ($\lambda_i \geq 0$), the maximum step size $\alpha_{\text{max}}$ for \eqref{eq:delta_y} is determined as:
\begin{align}
\alpha_{\text{max}} &= \operatorname{sup} \{  \alpha \in [0, 1]\  | \  \lambda + \alpha \Delta \lambda \succeq 0  \} \notag \\
&= \min \{1, \min\{-\lambda_i / \Delta \lambda_i | \Delta \lambda_i < 0\}\}.
\label{eq:a_max}
\end{align}

\noindent With \eqref{eq:a_max}, a backtracking line search begins with $\alpha = 0.99 \alpha_{\text{max}}$ and iteratively reduces $\alpha$ by a factor $\zeta \in (0,1)$ until the following condition is met:
\begin{equation}
\begin{aligned}
\big \Vert r_{\text{KKT}}(s_t^{+}, \lambda^{+}) \Vert \leq (1 - \kappa \alpha) \Vert r_{\text{KKT}}(s_t, \lambda) \Vert + \epsilon_{\text{tol}},
\label{eq:backtracking}
\end{aligned}
\end{equation}

\noindent where $\kappa \in (0,1)$ and $\epsilon_{\text{tol}} > 0$ is a tolerance parameter accounting for data noise. The updates used in \eqref{eq:backtracking} are:
\begin{align}
s_t^{+} = s_t + \alpha \Delta s_t, \ \lambda^{+} = \lambda + \alpha \Delta \lambda. \notag
\end{align}

\noindent The surrogate duality gap is computed as:
\begin{equation}
\begin{aligned}
\hat{\delta}(s_t,\lambda) = -\phi(s_t)^{\top} \lambda.
\end{aligned}
\label{eq:gap}
\end{equation}

\noindent Equation \eqref{eq:gap} indicates convergence: as $\hat{\delta}(s_t,\lambda)$ decreases towards zero, the iterates approach primal and dual feasibility. The optimization terminates when $\Vert r_{\text{dual}} \Vert$ and $\hat{\delta}(s_t,\lambda)$ falls below specified tolerances. The overall primal-dual interior-point method is summarized in Algorithm \ref{alg:primal_dual}.

\begin{algorithm}
\DontPrintSemicolon
\KwInput{Feasible $s_t$, $\lambda > 0$, $\mu > 1$, $\epsilon_{\text{feas}} > 0$, $\epsilon_{\text{gap}} > 0$}
\KwOutput{Optimized $s_{t}$ and $\lambda$}
\Repeat{\upshape $\Vert r_{\text{dual}} \Vert \leq \epsilon_{\text{feas}}$ \upshape\textbf{and} $\hat{\delta}(s_t,\lambda) \leq \epsilon_{\text{gap}}$}{
  Set barrier parameter: $\beta \leftarrow \mu m / \hat{\delta}(s_t,\lambda)$\;
  Compute the primal-dual direction: $\Delta y$\;
  Determine step size $\alpha$ via line search\;
  Update variables: $y \leftarrow y + \alpha \Delta y$\;
}%
\caption{Primal-Dual Interior-Point Method}
\label{alg:primal_dual}
\end{algorithm}

\subsection{Iteratively Reweighted Least Squares (IRLS)}

To enhance robustness, the interior-point method is embedded in an outer IRLS loop. At each iteration, residuals are scored with a MAD-based robust z-score to down-weight outliers and discard extreme segments. The reweighted problem is then solved by the inner primal-dual interior-point method (i.e., Algorithm \ref{alg:primal_dual}), with convergence achieved in about three outer iterations. The overall sliding-window IRLS estimator for quadrotor motor efficiency is summarized in Algorithm \ref{alg:irls}.
\begin{algorithm}
\DontPrintSemicolon
\KwInput{Feasible $s^{(0)}_t$, Max iterations $N_{\text{IRLS}}$}
\KwOutput{Optimized $s_{t}$}
\For{$k = 1$ \KwTo $N_{\text{IRLS}}$}{
  Compute trajectory residuals $r_i(s^{(k-1)}_t)$\;
  Compute weighted residual energy $e_i$\;
  Compute robust z-score $z_i$\;
  Update weights $w_i$ and hard-reject if $z_i > z_\text{hard}$\;
  Form full block-diagonal weight matrix $G$\;
  Run primal-dual interior-point method for $s^{(k)}_{t}$\;
}
\Return{$s^{(k)}_{t}$}
\caption{Overall IRLS Estimator}
\label{alg:irls}
\end{algorithm}
 
\subsection{Derivatives of Trajectory Residuals}

This section derives the partial derivatives of all residuals with respect to the motor efficiency parameters $s$. The velocity residual derivative with respect to $s$ is:
\begin{equation}
\begin{aligned}
\frac{\partial r_{v,t}(s)}{\partial s} =
\frac{\Delta t}{m}
\begin{bmatrix}
f_1 r_{13} & f_2 r_{13} & f_3 r_{13} & f_4 r_{13} \\
f_1 r_{23} & f_2 r_{23} & f_3 r_{23} & f_4 r_{23} \\
f_1 r_{33} & f_2 r_{33} & f_3 r_{33} & f_4 r_{33} \\
\end{bmatrix}.
\end{aligned}
\label{eq:v_derivative}
\end{equation}

\noindent The position residual derivative with respect to $s$ is:
\begin{equation}
\begin{aligned}
\frac{\partial r_{x,t}(s)}{\partial s} =
\frac{\Delta t^2}{2m}
\begin{bmatrix}
f_1 r_{13} & f_2 r_{13} & f_3 r_{13} & f_4 r_{13} \\
f_1 r_{23} & f_2 r_{23} & f_3 r_{23} & f_4 r_{23} \\
f_1 r_{33} & f_2 r_{33} & f_3 r_{33} & f_4 r_{33} \\
\end{bmatrix}.
\end{aligned}
\label{eq:x_derivative}
\end{equation}

\noindent The angular velocity residual derivative with respect to $s$ is:
\begin{equation}
\renewcommand{\arraystretch}{1.4}
\begin{aligned}
\frac{\partial r_{\Omega,t}(s)}{\partial s} =
\Delta t
\begin{bmatrix}
\frac{f_1 d}{J_{xx}} & \frac{-f_2 d}{J_{xx}} & \frac{-f_3 d}{J_{xx}} & \frac{f_4 d}{J_{xx}} \\
\frac{-f_1 d}{J_{yy}} & \frac{-f_2 d}{J_{yy}} & \frac{f_3 d}{J_{yy}} & \frac{f_4 d}{J_{yy}} \\
\frac{f_1 c_{\tau f}}{J_{zz}} & \frac{-f_2 c_{\tau f}}{J_{zz}} & \frac{f_3 c_{\tau f}}{J_{zz}} & \frac{-f_4 c_{\tau f}}{J_{zz}} \\
\end{bmatrix}.
\end{aligned}
\label{eq:W_derivative}
\end{equation}

\noindent Recall \eqref{eq:deltaR_measured} and define the rotation matrix difference:
\begin{equation}
\begin{aligned}
\delta R^{\top} \coloneqq
\begin{bmatrix}
\delta r_{11} & \delta r_{21} & \delta r_{31} \\
\delta r_{12} & \delta r_{22} & \delta r_{32} \\
\delta r_{13} & \delta r_{23} & \delta r_{33} \\
\end{bmatrix}.
\end{aligned}
\label{eq:delta_R_t}
\end{equation}

\noindent Based on \eqref{eq:delta_R_t}, the rotation residual derivative with respect to $s$ is:
\begin{equation}
\begin{aligned}
&\frac{\partial r_{R,t}(s)}{\partial s} =\\
& \frac{\Delta t^2}{2}
\begin{bmatrix}
-\frac{(\delta r_{23} - \delta r_{32}) f_1 d}{J_{xx}} + \frac{(\delta r_{31} - \delta r_{13}) f_1 d}{J_{yy}} - \frac{(\delta r_{12} - \delta r_{21}) f_1 c_{\tau f}}{J_{zz}} \\
\frac{(\delta r_{23} - \delta r_{32}) f_2 d}{J_{xx}} + \frac{(\delta r_{31} - \delta r_{13}) f_2 d}{J_{yy}} + \frac{(\delta r_{12} - \delta r_{21}) f_2 c_{\tau f}}{J_{zz}} \\
\frac{(\delta r_{23} - \delta r_{32}) f_3 d}{J_{xx}} - \frac{(\delta r_{31} - \delta r_{13}) f_3 d}{J_{yy}} - \frac{(\delta r_{12} - \delta r_{21}) f_3 c_{\tau f}}{J_{zz}} \\
-\frac{(\delta r_{23} - \delta r_{32}) f_4 d}{J_{xx}} - \frac{(\delta r_{31} - \delta r_{13}) f_4 d}{J_{yy}} + \frac{(\delta r_{12} - \delta r_{21}) f_4 c_{\tau f}}{J_{zz}}
\end{bmatrix}.
\end{aligned}
\label{eq:R_derivative}
\end{equation}

\noindent With the analytical expressions derived in \eqref{eq:v_derivative}--\eqref{eq:W_derivative} and \eqref{eq:R_derivative}, the Jacobian $J_r(s)$ in \eqref{eq:Jr} can be constructed to compute the Newton step in the primal-dual interior-point method.

\subsection{EKF Baseline}

For comparison, we implemented an extended Kalman filter (EKF) as a baseline. 
The EKF state is defined as
\begin{equation}
s_{\mathrm{EKF}} =
\begin{bmatrix}
x^{\top} & v^{\top} & \Omega^{\top} & \mathrm{vec}(R)^{\top} & \eta^{\top}
\end{bmatrix}^{\top} \in \mathbb{R}^{22}, \notag
\end{equation}

\noindent where the dynamics of motor efficiency are modeled as random walks:
\begin{equation}
\dot{\eta} = w_\eta, \quad w_\eta \sim \mathcal{N}(0, Q_\eta). \notag
\end{equation}

\noindent This EKF serves as a filter-based baseline for comparison against the proposed batch optimization approach. No explicit outlier rejection is applied in the EKF baseline.

\section{Simulations}
\label{sec:simulation}

This section presents simulation results to validate the effectiveness of the proposed motor efficiency estimation framework under various degradation and fault scenarios.  The MATLAB code used for the simulations can be accessed at \href{https://github.com/shengwen-tw/irls-quadrotor-motor-efficiency-estimator}{https://github.com/shengwen-tw/irls-quadrotor-motor-efficiency-estimator}.

\subsection{Simulation Setup}

The quadrotor parameters and controller gains are configured to resemble the characteristics of a typical F450 platform, as summarized in Table \ref{tab:quadrotor_params}.

\begin{table}[H]
\centering
\caption{Quadrotor Parameters and Control Gains}
\begin{tabular}{|c|c|}
\hline
\textbf{Parameter} & \textbf{Value} \\
\hline
\(m\) & 1.0 kg \\
\(J\) & $\operatorname{diag}(0.01466, 0.01466, 0.02848)\ \text{kg} \cdot \text{m}^2$ \\
\(d\) & 0.225 m \\
\(c_{\tau f}\) & $0.009012\ \text{m}$ \\
\hline
$k_x$ & $\operatorname{diag}(9,9,12)$ \\
$k_v$ & $\operatorname{diag}(7,7,12)$ \\
$k_R$ & $\operatorname{diag}(10,10,10)$ \\
$k_\Omega$ & $\operatorname{diag}(2,2,2)$ \\
\hline
\end{tabular}
\label{tab:quadrotor_params}
\end{table}

\noindent The desired trajectory consists of a circular motion in the $xy$-plane combined with a constant value in the $z$-direction, defined as:
\begin{equation}
\begin{aligned}
x_d(t) =
\begin{bmatrix}
3 \cos(0.2\pi t) \\
3 \sin(0.2 \pi t) \\
-1
\end{bmatrix},\ 
v_d(t) =
\begin{bmatrix}
-0.6 \pi \sin(0.2 \pi t) \\
0.6 \pi \cos(0.2 \pi t) \\
0
\end{bmatrix}.
\notag
\end{aligned}
\end{equation}

\noindent The desired body x-axis direction and angular velocity are given by:
\begin{equation}
\begin{aligned}
b_{1d}(t) =
\begin{bmatrix}
\cos(0.1 \pi t) \\
\sin(0.1 \pi t) \\
0
\end{bmatrix},\ 
\Omega_d(t) =
\begin{bmatrix}
0 \\ 0 \\ 0
\end{bmatrix}.
\notag
\end{aligned}
\end{equation}

\noindent To evaluate the robustness of the proposed estimator, the following conditions are simulated:\newline

\noindent $\bullet$ \textbf{Voltage-Induced Gradual Degradation}
\newline Motor efficiency $\eta_i(t)$ degrades exponentially with battery voltage $V(t)\in \mathbb{R}$:
\begin{align}
\eta_i(t) = \eta_i(0) \exp(\xi (V(t) - V(0))),\ i = 1,\cdots,4, \notag
\end{align}

\noindent where $\eta_i(0)$ is the initial efficiency and $\xi > 0$ is a voltage sensitivity constant.

\noindent $\bullet$ \textbf{Abrupt Fault Injection}
\newline At specified time intervals, selected motors experience sudden efficiency drops:
\begin{equation}
\begin{aligned}
\eta_i(t) = 
\begin{cases} 
 0.5       & \text{if\ } t \in \text{fault interval for motor } i \\
 \eta_i(t) & \text{otherwise}.
\end{cases} \notag
\end{aligned}
\end{equation}

\noindent $\bullet$ \textbf{Random Thrust Noise}
\newline To emulate real-world disturbances, random noise is added to the motor thrust vector:
\begin{align}
f_{i,\epsilon_f} = f_i \cdot \exp(\epsilon_f),\ \epsilon_f \sim \mathcal{N}(0, \sigma_f), \notag
\end{align}
where $\sigma_f$ denotes the thrust noise level.

\subsection{Simulation Results and Transition Robustness}

\begin{figure}[H]
    \centering
    \includegraphics[width=0.95\linewidth]{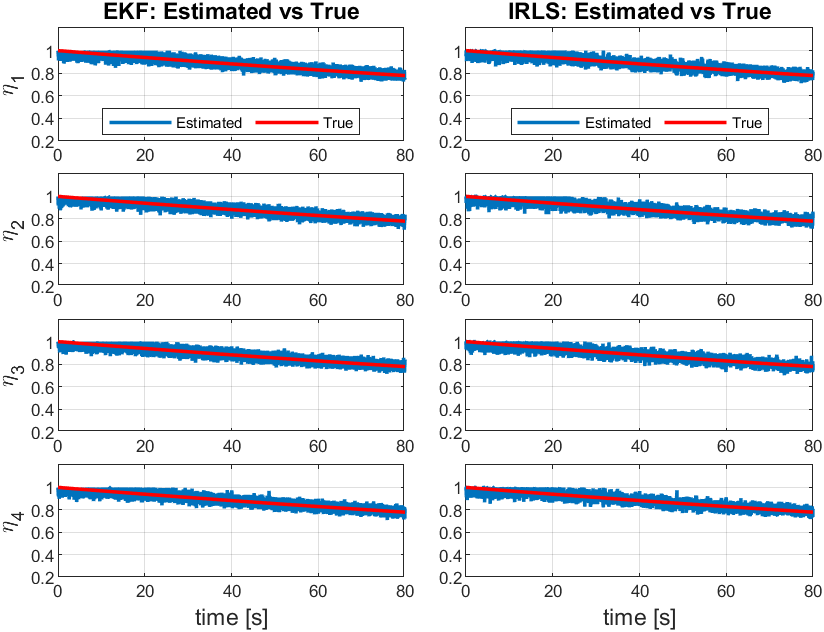}
    \caption{Comparison between the EKF baseline and the proposed estimator under gradual voltage-induced degradation with thrust noise.}
    \label{fig:degrade}
\end{figure}

\begin{figure}[H]
    \centering
    \includegraphics[width=0.95\linewidth]{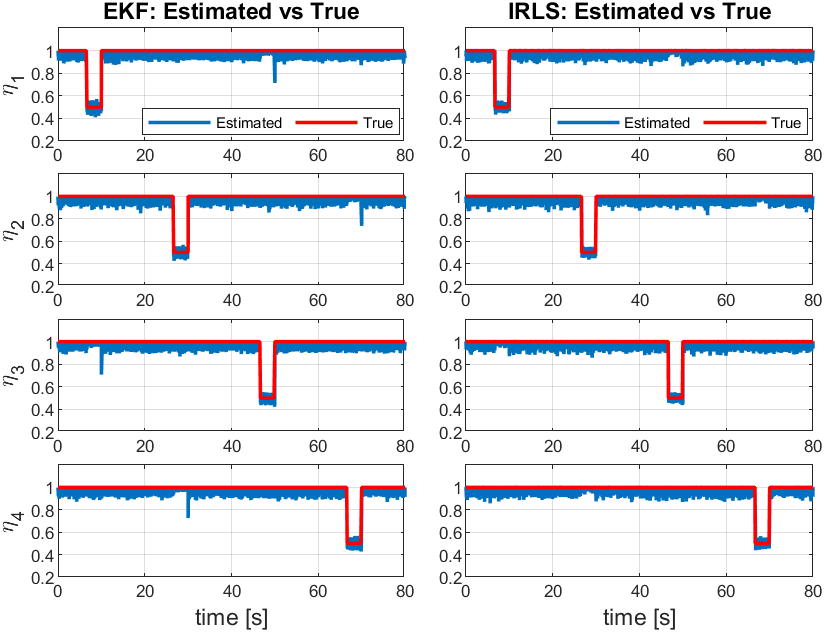}
    \caption{Comparison between the EKF baseline and the proposed estimator under abrupt motor fault injection with thrust noise.}
    \label{fig:fault_injection}
\end{figure}

\begin{figure}[H]
    \centering
    \includegraphics[width=0.95\linewidth]{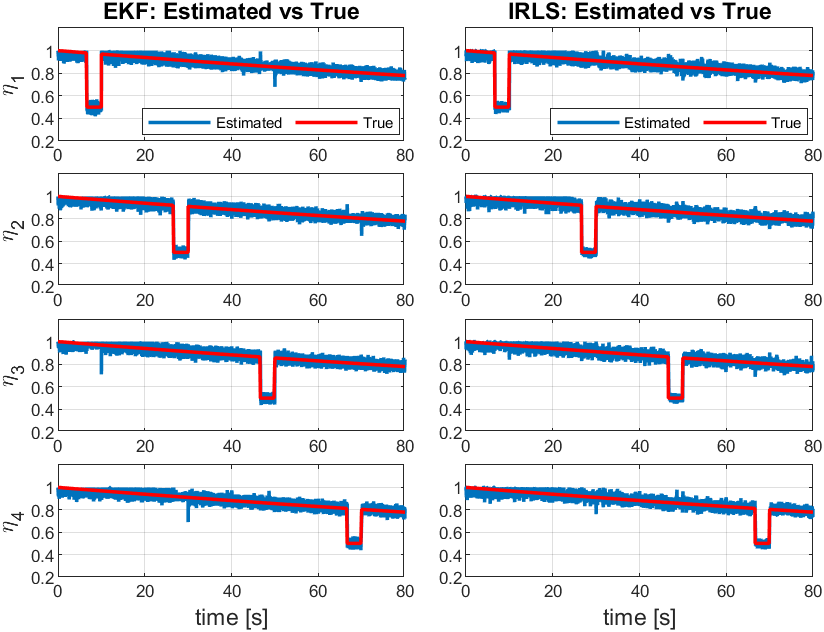}
    \caption{Comparison between the EKF baseline and the proposed estimator under combined degradation and abrupt fault scenarios with thrust noise.}
    \label{fig:combined_test}
\end{figure}

Fig. \ref{fig:degrade}-\ref{fig:combined_test} compare the EKF baseline and the proposed estimator under gradual degradation, abrupt fault injection, and combined cases, all with thrust noise level $\sigma_f = 0.07$. While both methods achieve similar accuracy when calibrated to the same root mean square error (RMSE) and standard deviation levels, their behavior during transitions differs markedly. The EKF exhibits pronounced spikes whenever abrupt changes occur, as incoherent measurements are directly incorporated into the filter updates. In contrast, the proposed IRLS-based estimator effectively down-weights outliers and maintains smooth estimates through transitions, accurately reflecting the true motor efficiency even under sudden faults.

\subsection{RMSE and Spike Comparison Across Scenarios}

\begin{figure}[H]
    \centering
    \includegraphics[width=0.9\linewidth]{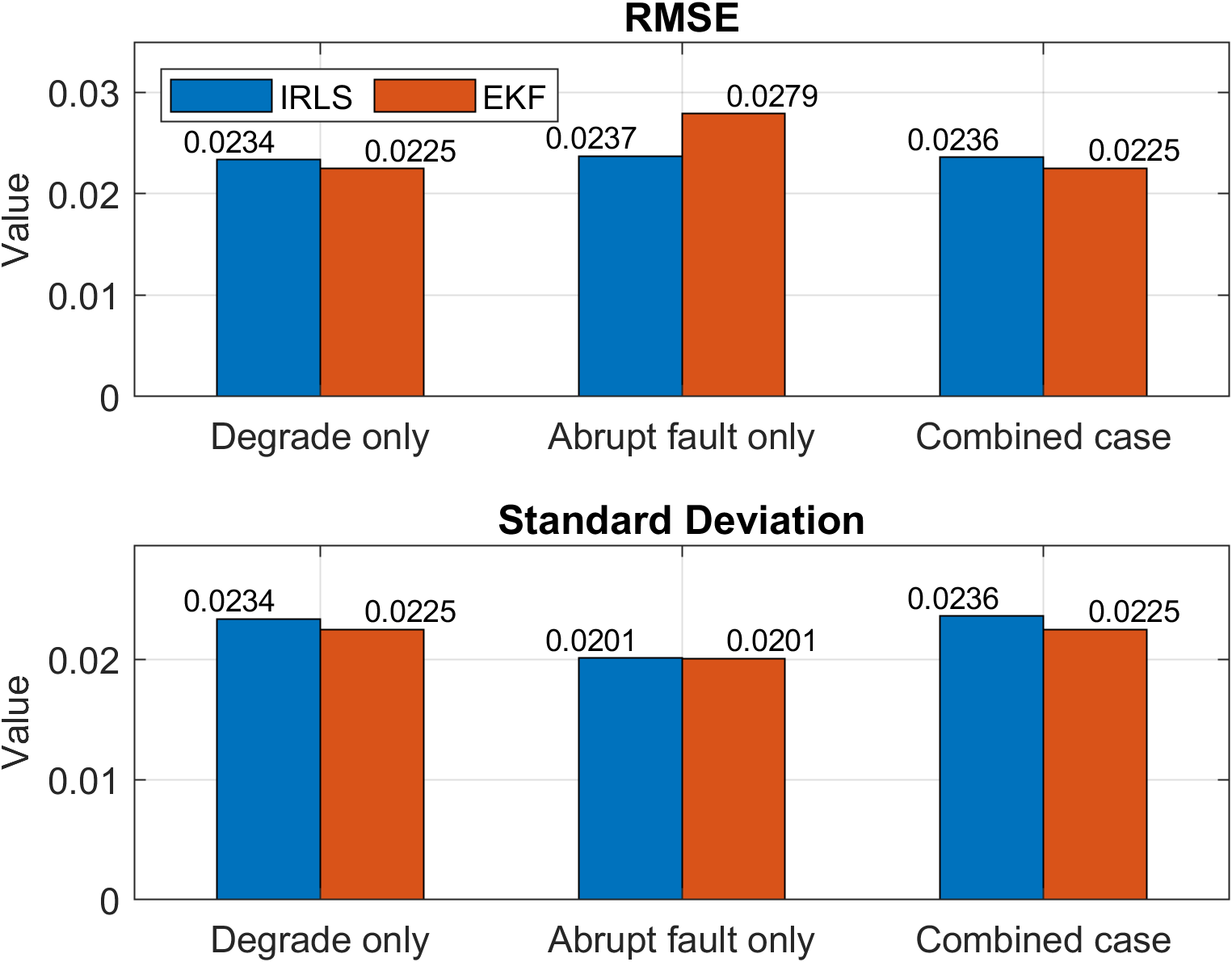}
    \caption{Root mean square error (RMSE) and standard deviation of motor efficiency estimation under various degradation and fault scenarios, used to show that the EKF baseline and the proposed method are calibrated to comparable average accuracy.}
    \label{fig:rmse}
\end{figure}

\begin{figure}[H]
    \centering
    \includegraphics[width=0.9\linewidth]{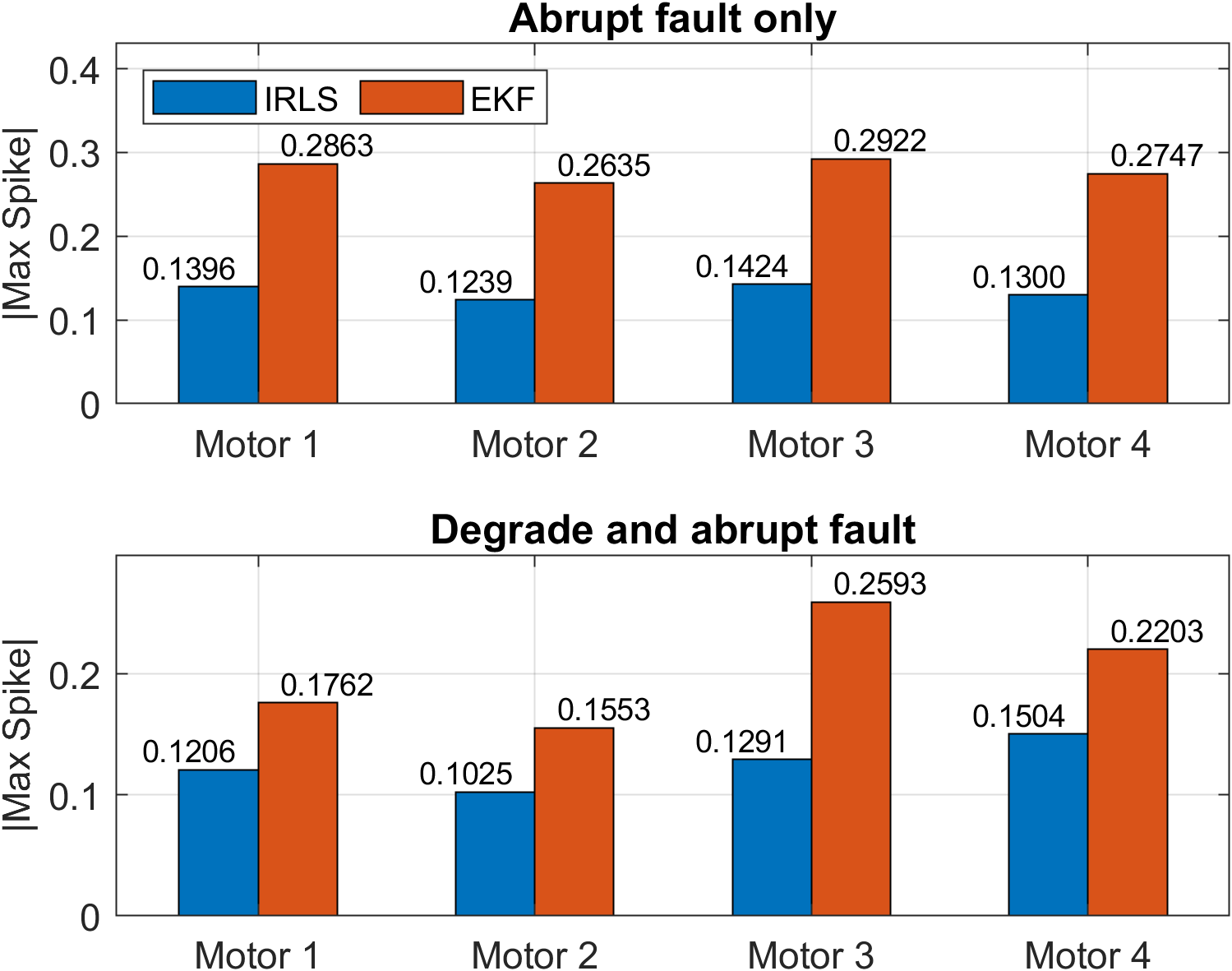}
    \caption{Comparison of maximum estimation spikes for each motor between the EKF baseline and the proposed method under (a) abrupt-only faults and (b) combined degradation-and-abrupt faults.}
    \label{fig:spike}
\end{figure}

Fig. \ref{fig:rmse} summarizes the RMSE and standard deviation across multiple scenarios, with both methods calibrated to the same accuracy level. Under this normalization, Fig. \ref{fig:spike} highlights the main difference: the proposed method consistently suppresses the transient spikes observed in the EKF during fault injection and combined scenarios. These results confirm that the proposed approach offers robustness advantages while preserving comparable average accuracy.

\subsection{Estimation and Optimization Convergence}

\begin{figure}[H]
    \centering
    \includegraphics[width=0.9\linewidth]{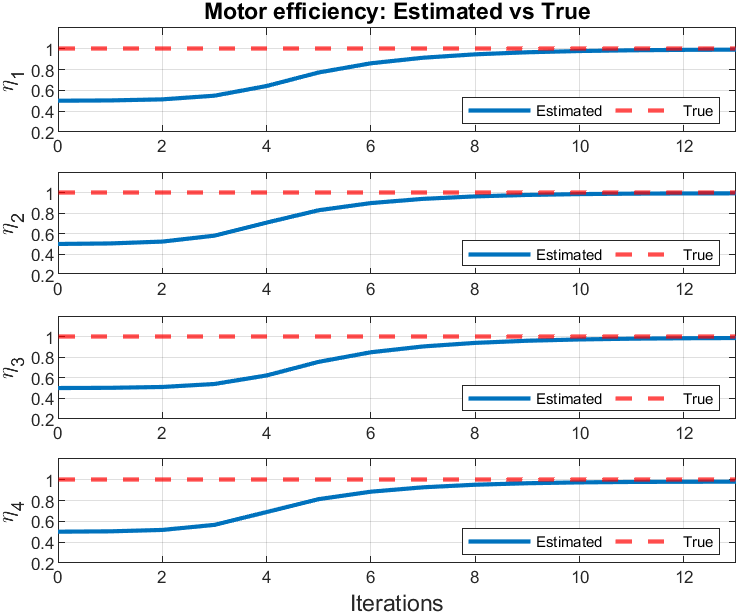}
    \caption{Convergence of motor efficiency estimates from an initial guess of $0.5$ over one full loop of the interior-point method.}
    \label{fig:efficiency_iteration}
\end{figure}

\begin{figure}[H]
    \centering
    \includegraphics[width=0.95\linewidth]{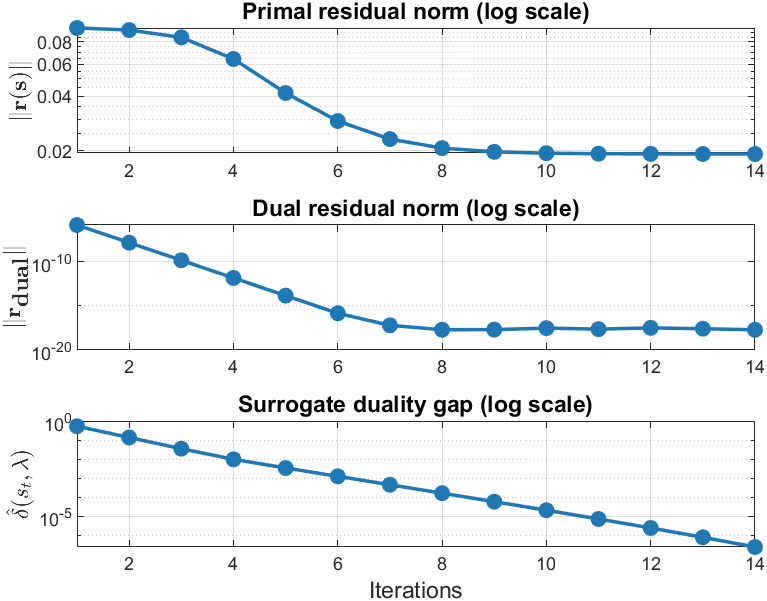}
    \caption{Evolution of primal residual, dual residual, and surrogate duality gap over one full loop of the interior-point method, confirming KKT convergence.}
    \label{fig:kkt_iteration}
\end{figure}

Fig. \ref{fig:efficiency_iteration} and Fig. \ref{fig:kkt_iteration} illustrate the convergence properties of the proposed estimator. Fig. \ref{fig:efficiency_iteration} shows that starting from an initial guess of $0.5$, the efficiency estimates converge to the true values within one full loop of the interior-point method. Fig. \ref{fig:kkt_iteration} further presents the evolution of primal residual, dual residual, and surrogate duality gap, all of which decrease monotonically within the same loop. Together, these results demonstrate that the optimization reliably satisfies KKT optimality conditions and converges efficiently.

\section{Conclusion}
\label{sec:conclusion}

This paper presented a data-driven framework for online estimation of quadrotor motor efficiency based on residual minimization. The estimation problem was formulated as a constrained nonlinear optimization within a sliding-window structure, solved efficiently using a primal-dual interior-point method. To enhance robustness, an outer iteratively reweighted least squares (IRLS) loop with robust z-score weighting was integrated, enabling rejection of incoherent measurements and suppression of transient spikes.

Compared to traditional filter-based methods such as EKF, the proposed batch-mode optimization achieves similar accuracy while exhibiting markedly improved robustness during abrupt transitions. Simulation results under degradation, fault injection, and combined scenarios showed that the estimator avoids the large spikes observed in the EKF, particularly in motor clipping cases, by effectively rejecting outliers.

The methodology is well-suited for applications in fault detection and isolation, health monitoring, and predictive maintenance in aerial robotic systems. Its formulation supports extensibility toward learning-enabled estimation and control architectures. Future research may focus on enhancing robustness against unmodeled effects such as wind disturbances and external perturbations.


\printbibliography

\end{document}